\documentclass[10pt,conference,compsocconf]{IEEEtran}
\usepackage{times}

\usepackage{caption}
\usepackage{cite}
\usepackage{url}
\usepackage{epsfig}
\usepackage{subfigure}
\usepackage{pifont}
\usepackage{amssymb}
\usepackage{amstext}
\usepackage{amsmath}
\usepackage{amsthm}
\usepackage{multicol}
\usepackage[latin1]{inputenc}

\begin{document}
\pagenumbering{gobble}
\title{\textbf{\Large About Microservices, Containers \\
and their Underestimated Impact on Network Performance}}


\author{\IEEEauthorblockN{~\\[-0.4ex]\large Nane Kratzke\\[0.3ex]\normalsize}
\IEEEauthorblockA{Lübeck University of Applied Sciences, Center of Excellence CoSA\\
Lübeck, Germany\\
email: {\tt nane.kratzke@fh-luebeck.de}}}

\maketitle

\begin{abstract}
Microservices are used to build complex applications composed of small, independent and highly decoupled processes.  Recently, microservices are often mentioned in one breath with container technologies like Docker. That is why operating system virtualization experiences a renaissance in cloud computing. These approaches shall provide horizontally scalable, easily deployable systems and a high-performance alternative to hypervisors. Nevertheless, performance impacts of containers on top of hypervisors are hardly investigated. Furthermore, microservice frameworks often come along with software defined networks. This contribution presents benchmark results to quantify the impacts of container, software defined networking and encryption on network performance. Even containers, although postulated to be lightweight,  show a noteworthy impact to network performance. These impacts can be minimized on several system layers. Some design recommendations for cloud deployed systems following the microservice architecture pattern are derived.
\end{abstract}

\begin{IEEEkeywords}
Microservice; Container; Docker; Software Defined Network; Performance%
\end{IEEEkeywords}

\section{\uppercase{Introduction}}
\label{sec:introduction}
Microservices are applied by companies like Amazon, Netflix, or SoundCloud \cite{Fowler2014}\cite{Newman2015}. This architecture pattern is used to build big, complex and horizontally scalable applications composed of small, independent and highly decoupled processes communicating with each other using language-agnostic application programming interfaces (API). Microservice approaches and container-based operating system virtualization experience a renaissance in cloud computing. Especially container-based virtualization approaches are often mentioned to be a high-performance alternative to hypervisors \cite{SPF+2007}. \emph{Docker} \cite{docker} is such a container solution, and it is based on operating system virtualization using Linux containers. Recent performance studies show only little performance impacts to processing, memory, network or I/O \cite{FFRR2014}. That is why \emph{Docker} proclaims itself a "lightweight virtualization platform" providing a standard runtime, image format, and build system for Linux containers deployable to any Infrastructure as a Service (IaaS) environment.

  This study investigated the performance impact of Linux containers on top of hypervisor based virtual machines logically connected by an (encrypted) overlay network. This is a common use case in IaaS Cloud Computing being applied by popular microservice platforms like Mesos \cite{mesos}, CoreOS \cite{coreos} or Kubernetes \cite{kubernetes} (the reader may want to study a detailed analysis of such kind of platforms \cite{Kra2014b}). Nevertheless, corresponding performance impacts have been hardly investigated so far. Distributed cloud based microservice systems of typical complexity often use hypertext transfer protocol (HTTP) based and representational state transfer (REST) styled protocols to enable horizontally scalable system designs \cite{Fielding2000}. If these systems are deployed in public clouds, additional requirements for encrypted data transfer arise. There exist several open source projects providing such a microservice approach on top of IaaS provider specific infrastructures using this approach (e.g.
    Mesos,
    Kubernetes,
    CoreOS
     and more). These approaches are intended to be deployable to public or private IaaS infrastructures \cite{Kra2014b}. So in fact, these approaches apply operating system virtualization (containers) on top of hypervisors (IaaS infrastructures). Although    
almost all of these microservice frameworks rely heavily on the combination of containerization on top of hypervisors, and some of these approaches introduce additional overlay networking and data encryption layers, corresponding performance impacts have been hardly analyzed so far. Most performance studies compare container performance with virtual machine performance but not container performance on top of virtual machines (see Felter at al. \cite{FFRR2014} for a typical performance study). 

Because overlay networks are often reduced to distributed hashtable (DHT) or peer-to-peer approaches, this paper uses the term software defined virtual networks (SDVN). SDVNs, in the understanding of this paper, are used to provide a logical internet protocol (IP) network for containers on top of IaaS infrastructures.

Section \ref{sec:relatedwork} presents related work about state-of-the-art container approaches and SDVN solutions. Section \ref{sec:experimentdesign} explains the experiment design to identify performance impacts of containers, SDVNs and encryption. The benchmark tooling \cite{ping-pong} and the performance data collected is provided online \cite{sdvn-impact-database}. Resulting performance impacts are discussed in Section \ref{sec:discussion}. Derived design recommendations to minimize performance impacts on application,  overlay network and IaaS infrastructure layer are presented in concluding Section \ref{sec:conclusion}.

\section{\uppercase{Related Work}}
\label{sec:relatedwork}

\begin{figure*}[tb]
  \centering
   \subfigure[{Experiment to identify reference performance (\ding{110})}]{\epsfig{file = 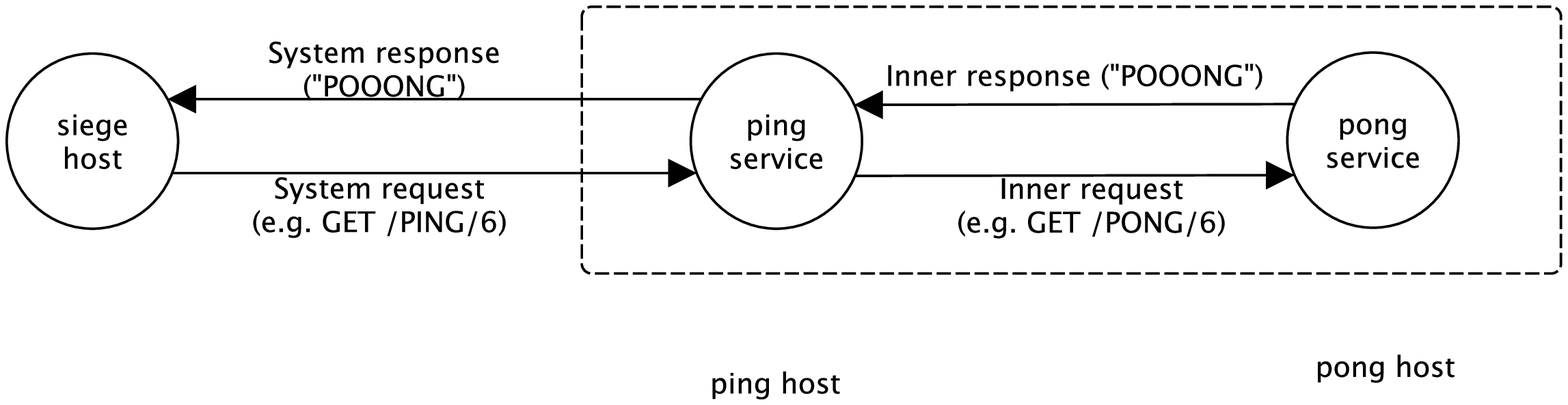, width = \columnwidth}}
   \subfigure[{Experiment to identify impact of Docker (\ding{108})}]{\epsfig{file = 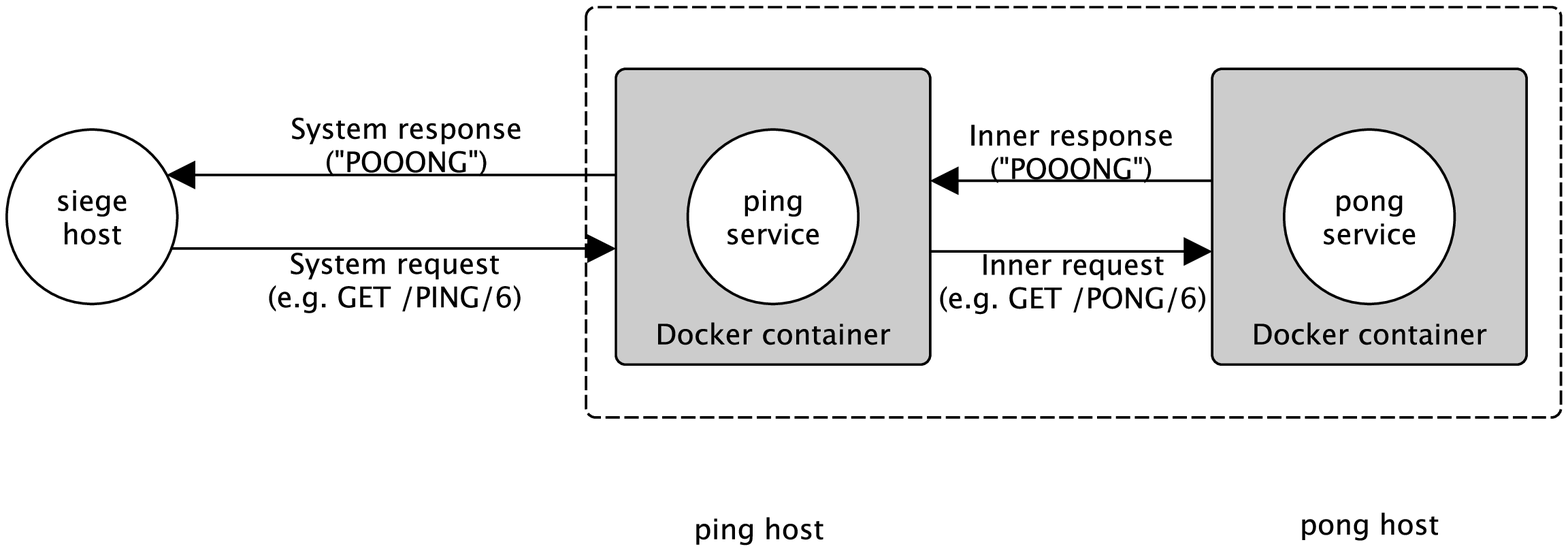, width = \columnwidth}}
   \subfigure[{Experiment to identify impact of SDVN (\ding{115})}]{\epsfig{file = 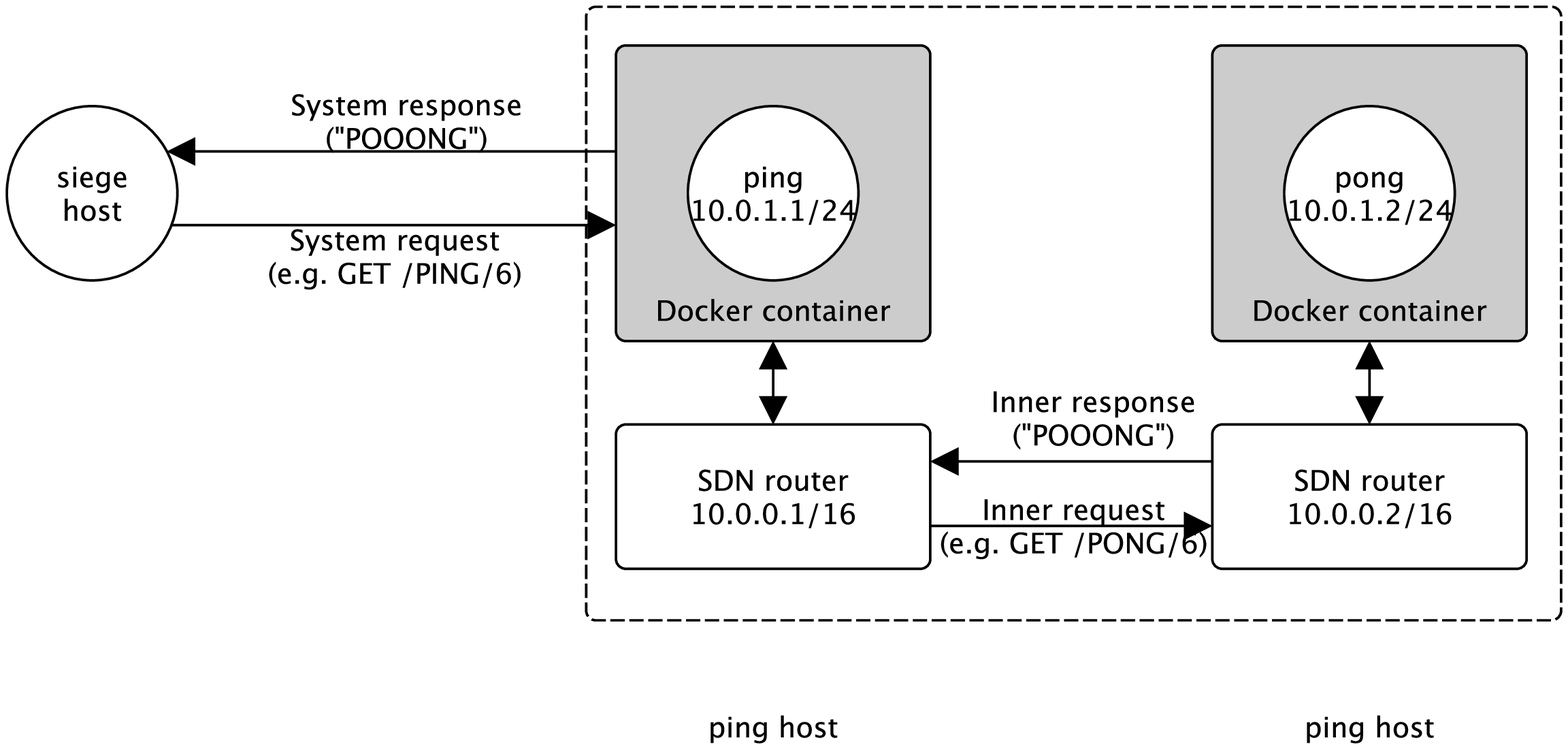, width = \columnwidth}}
   \subfigure[{Experiment to identify impact of encryption (\ding{116})}]{\epsfig{file = 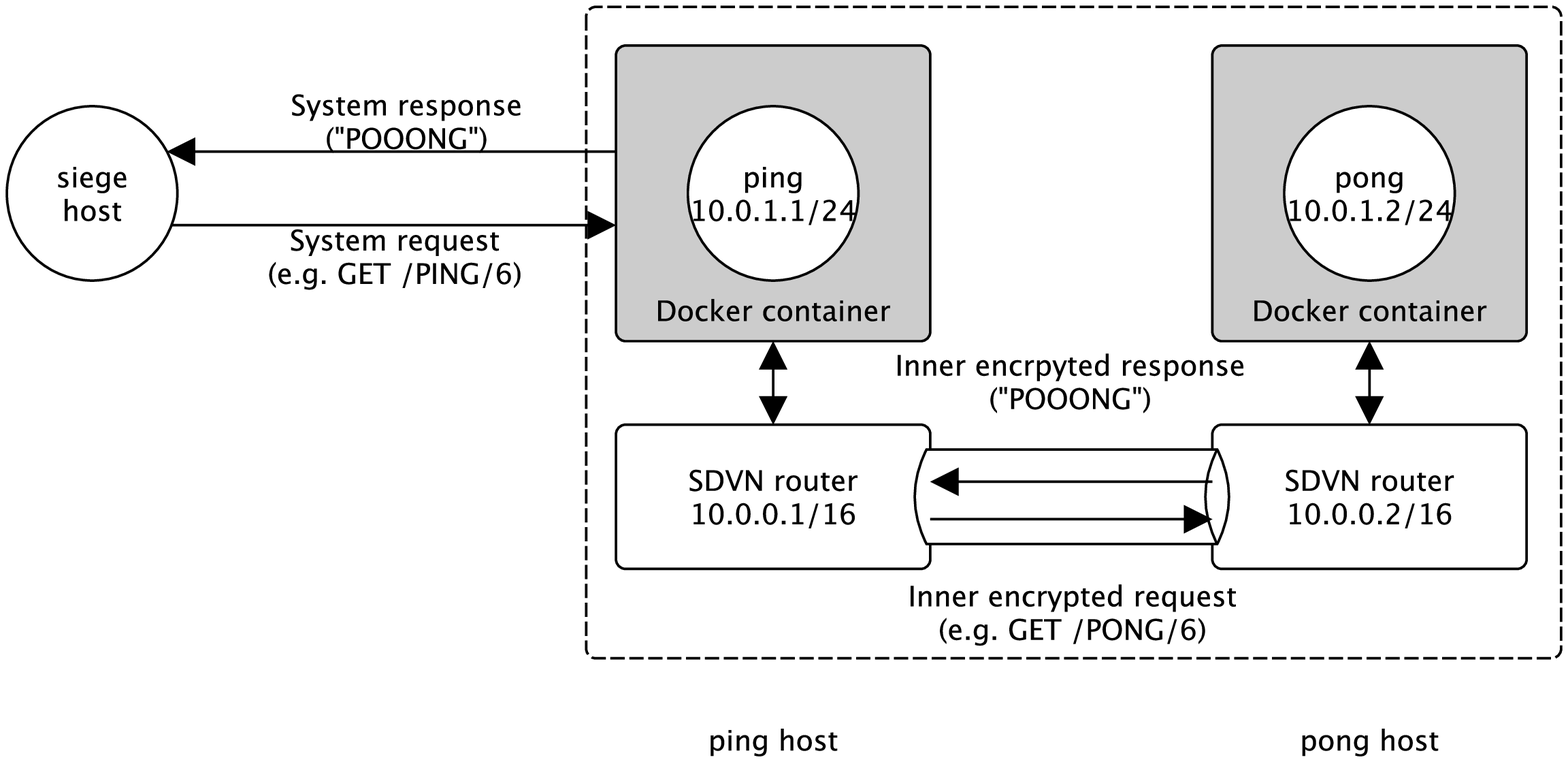, width = \columnwidth}}
  \caption{{Experiments}}
  \label{fig:reference-experiment}
\end{figure*}

Although container based operating system virtualization is postulated to be a scalable and high-performance alternative to hypervisors, hypervisors are the standard approach for IaaS cloud computing   \cite{SPF+2007}. Felter et al. provided a very detailed analysis on CPU, memory, storage and networking resources to explore the performance of traditional virtual machine deployments, and contrast them with the use of Linux containers provided via \emph{Docker}  \cite{FFRR2014}. Their results indicate that benchmarks that have been run in a \emph{Docker} container, show almost the same performance (floating point processing, memory transfers, network bandwidth and latencies, block I/O and database performances) like benchmarks run on "bare metal" systems. Nevertheless, Felter et al. did not analyze the impact of containers on top of hypervisors.

  Although there exist several SDVN solutions for \emph{Docker}, only one open source based SDVN has been identified, which is able to encrypt underlying data transfers: \emph{Weave} \cite{weave}. That is why other SDVN approaches for \emph{Docker} like \emph{flannel} \cite{flannel} or \emph{docknet} \cite{docknet} are not covered by this study. Pure virtual local area network (VLAN) solutions like \emph{Open vSwitch (OVS)} \cite{ovs} are not considered, because \emph{OVS} is not to be designed for operating system virtualization. So, \emph{weave} remained as the only appropriate SDVN candidate for this study. But the author is confident that this will change in the future and more encryptable SDVN solutions will arise.
  
  \emph{Weave} creates a network bridge on \emph{Docker} hosts to enable SDVN for \emph{Docker} containers. Each container on a host is connected to that bridge. A \emph{weave} router captures Ethernet packets from its bridge-connected interface in promiscuous mode. Captured packets are forwarded over the user datagram protocol (UDP) to \emph{weave} router peers running on other hosts. These UDP "connections" are duplex, can traverse firewalls and can be encrypted.
  
  To analyze the performance impact of containers, software defined networks and encryption, this paper considered several contributions on cloud related network performance analysis  (see\cite{MJ1998}, \cite{MMH2001}, \cite{VGNV2005}, \cite{JRM+2010}, \cite{WN2010}, \cite{JML+2014}). But none of these contributions focused explicitly on horizontally scalable systems with HTTP-based and REST-like protocols. To address this common use case for microservice architectures, this paper proposes the following experiment design.

\section{\uppercase{Experiment Design}}
\label{sec:experimentdesign}

\begin{figure*}[tb]
	\centering
	\epsfig{file = 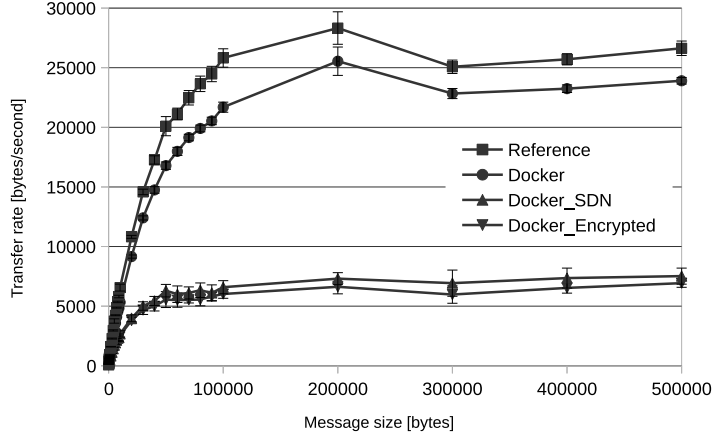, width = 0.55 \textwidth}
	\caption{{Absolute performance impact on transfer rates}}
	\label{fig:comparison}
\end{figure*}

  This study analyzed the network performance impact of container, SDVN and encryption layers on the performance impact of distributed cloud based systems using HTTP-based REST-based protocols. Therefore, five experiments have been designed (see Figure \ref{fig:reference-experiment}). The analyzed \emph{ping-pong} system relied on a REST-like and HTTP-based protocol to exchange data. Apachebench  \cite{apachebench} was used to collect performance data of the \emph{ping-pong} system. \emph{Siege}, \emph{ping} and \emph{pong} servers have been deployed to the Amazon Web Services (AWS) IaaS infrastructure on a m3.medium instance type. Experiments have been run in eu-west-1c availability zone (Ireland). The \emph{siege} server run the apachebench benchmark. The \emph{ping} and \emph{pong} application were developed using Googles Dart programming language \cite{dartlang}. To understand the performance impact of containers, SDVN and encryption to network performance, the study analyzed the data transfer rate $trans(m)$ of $m$ byte long messages.
  
\ding{110} The \textbf{reference experiment} shown in Figure \ref{fig:reference-experiment}(a), was used to collect reference performance data of the \emph{ping-pong} system deployed to different virtual machines interacting with a REST-like and HTTP based protocol. No containers, SDVN or encryption were used in this experiment. Further experiments added a container, a SDVN and an encryption layer to measure their impact on network performance. A \emph{ping} host interacts with a \emph{pong} host to provide its service. Whenever the \emph{ping} host is requested by \emph{siege} host, the \emph{ping} host relays the original request to the \emph{pong} host. The \emph{pong} host answers the request with a response message. The \emph{siege} host can define the inner message and system response message length by query. All requests are performed using the HTTP protocol. The \emph{siege} host is used to run several \emph{apachebench} benchmark runs with increasing requested message sizes to measure the system performance of the \emph{ping-pong} system. This paper refers to measured transfer rates for a message size of $m$ (bytes) of this experiment as $trans_\square(m)$. These absolute values are presented in Figure \ref{fig:comparison}.

\ding{108} The intent of the \textbf{Docker experiment} was to figure out the impact of an additional container layer to network performance (Figure  \ref{fig:reference-experiment}(b)). So, the \emph{ping} and \emph{pong} services are provided as containers to add an additional container layer to the reference experiment. Every performance impact must be due to this container layer. The measured transfer rates are denominated as $trans_\bigcirc(m)$. The impact of containers on transfer rates is calculated as follows and presented in Figure \ref{fig:relcomparison}(a):
\begin{equation}
\begin{split}
\bigcirc_{trans}(m)&=\frac{trans_\bigcirc(m)}{trans_\square(m)}
\end{split}
\end{equation}

\ding{115} The intent of the \textbf{SDVN experiment} shown in Figure \ref{fig:reference-experiment}(c) was to figure out the impact of an additional SDVN layer to network performance. This experiment connects \emph{ping} and \emph{pong} containers by a SDVN. So, every data transfer must pass the SDVN solution between \emph{ping} and \emph{pong}. This paper refers to measured transfer rates for a message size of $m$ (bytes) of this experiment as $trans_\Delta(m)$. The impact of SDVN on transfer rates for a message size $m$ is calculated as follows and presented in Figure \ref{fig:relcomparison}(a):
\begin{equation}
\begin{split}
\Delta{trans}(m)&=\frac{trans_\Delta(m)}{trans_\square(m)} \\
\end{split}
\end{equation}

\ding{116} The \textbf{encryption experiment} (see Figure \ref{fig:reference-experiment}(d)) figured out the impact of an additional data encryption layer on top of  a SDVN layer. Additionally, this experiment encrypts the SDVN network. Corresponding measured transfer rates are denominated as $trans_\nabla(m)$. The impact of SDVN on transfer rates for a message size $m$ is calculated as follows and is presented in Figure \ref{fig:relcomparison}(a):
\begin{equation}
\begin{split}
\nabla{trans}(m)&=\frac{trans_\nabla(m)}{trans_\square(m)} \\
\end{split}
\end{equation}

$\blacktriangleright\hspace{-0.1cm}\blacktriangleleft$ The intent of \textbf{cross-regional experiment} was to figure out the impact of an cross-regional deployment to network performance. Although this was not the main focus of the study, this use case has been taken into consideration to generate a more graspable performance impact understanding for the reader. The setting has been the same as in Figure \ref{fig:reference-experiment}(a), except that the \emph{ping} and \emph{pong} hosts were deployed to different regions of the AWS infrastructure. \emph{ping} (and the \emph{siege} host) were deployed to the AWS region eu-west-1c (EU, Ireland) and the \emph{pong} host was deployed to AWS region ap-northeast-1c (Japan, Tokyo). Data from this experiment is only used to compare container, SDVN and encrypted SDVN performance with a cross-regional performance impact in a qualitative manner. A cross-regional impact might be more intuitively graspable for the reader. The cross-regional impact on transfer rates is presented in Figure \ref{fig:relcomparison}(a) as a lightgrey line.

\section{\uppercase{Discussion of Results}}
\label{sec:discussion}

  The results presented are based on more than 12 hours of benchmark runs, which resulted in a transfer of over 316GB of data requested by more than 6 million HTTP requests (see Table \ref{tab:characteristics}). Reference and Docker experiments show only minor standard deviations. The maximum deviations were measured for very small message sizes (10 and 20 byte). Standard deviations increased with the introduction of SDVN. So, the collected deviation data indicates that the experiment setting produces reliable data (increasing deviations of SDVN experiment have to do with the technical impacts of SDVNs, they are not due to experiment design and will be discussed by this paper).

\begin{figure*}[tbh]
	\centering
	\subfigure[{comparison of transfer rates (100\% means no loss)}]{\epsfig{file = 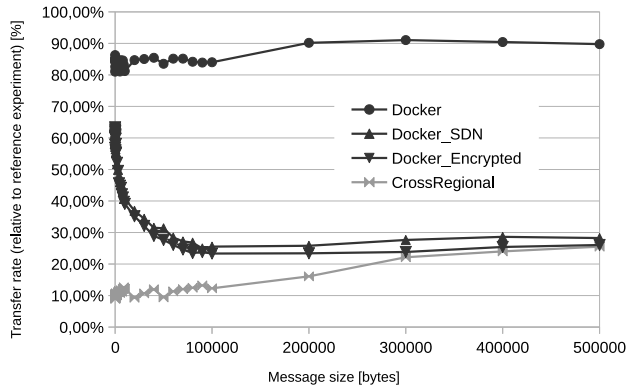, width = 1.0 \columnwidth}}
	\subfigure[{Resulting transfer losses (100\% means no loss)}]{\epsfig{file = 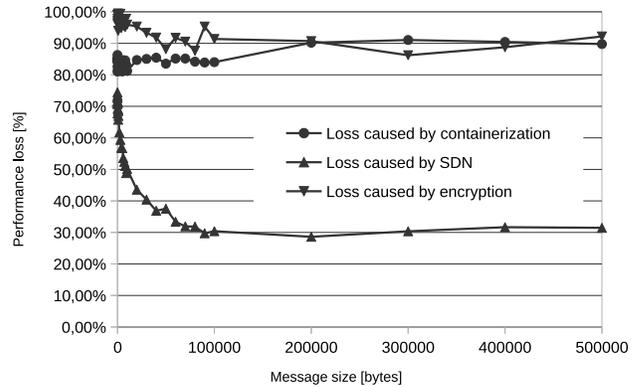, width = 1.0 \columnwidth}}
	\caption{{Relative comparison of performance indicators}}
	\label{fig:relcomparison}
\end{figure*}

\paragraph*{\ding{108} Container impact}

  Containers are stated to be lightweight and to have only negligible performance impacts \cite{FFRR2014}. The \emph{Docker} experiment shows a somewhat different picture. A non negligible performance loss can be identified for data transfer rates (see Figure \ref{fig:comparison}). An additional container layer on top of a bare virtual machine reduces the performance to 80\% (for message sizes smaller than 100 kBytes) to 90\% (for message sizes greater than 100kBytes). The study also included a cross-zone deployment (similar to the cross-region deployment but in two different availability zones of the same AWS region). It turned out that a cross-zone deployment shows almost the same performance like an one-zone deployment (reference experiment). Containers show a significant higher performance impact than cross-zone deployments in IaaS clouds. So, compared with cross-zone deployments (non-measurable effects), we have to say that containers have measurable (non-negligible) impacts to network performance.

\paragraph*{\ding{115} SDVN impact}

The impact of analyzed SDVN solution \emph{weave} reduces data transfer rates from 25 kB/s to about 7,5kB/s (see Figure \ref{fig:comparison}). Figure \ref{fig:relcomparison}a shows the relative performance of the SDVN experiment. SDVN experiments show only 60\% performance of the reference experiment for small message sizes going down to about 25\% performance for message sizes greater than 100kB.  The SDVN experiment shows a comparable performance impact like a cross-regional deployment (for big message sizes, Ireland $\leftrightarrow$ Japan). \emph{Weave} SDVN routers are provided as \emph{Docker} containers. The SDVN experiment measures the performance impact of containerization \textbf{and} SDVN. To identify the pure SDVN effect, the reader has to compare SDVN data ($\Delta$) relative to the performance data of containers ($\bigcirc$) to exclude the container effects (see Figure \ref{fig:relcomparison}b).  

\begin{equation}
\label{sdn_impact}
impact_\Delta(m)=\frac{trans_\Delta(m)}{trans_\bigcirc(m)}
\end{equation}

To avoid container losses, SDVN solutions should be provided directly on the host and not in a containerized form. This should reduce the performance loss about 10\% to 20\%.  Furthermore, it is noted that tests were running on a single-core virtual machine (m3.medium type). In saturated network load situations, the \emph{weave} router contends for CPU with the application processes, so it will saturate faster compared with Reference or \emph{Docker} experiment where the network is handled by the hypervisor and physical network, outside of such contention. This effect explains the severe performance impacts shown in Figure \ref{fig:relcomparison}. That lead us to the design conclusion that SDVN solutions should always run on multi-core systems to avoid severe performance impacts due to contention.

\paragraph*{\ding{116} Encryption impact}

  Additional encryption shows only minor impacts to transfer rates compared with SDVN without encryption (see Figure \ref{fig:comparison}). So, most of the performance losses are due to SDVN and not because of encryption. To identify the pure encryption effect, encrypted SDVN data ($\nabla$) has to be compared relative to the performance data of SDVN experiment ($\Delta$).   

\begin{equation}
impact_\nabla(m)=\frac{trans_\nabla(m)}{trans_\Delta(m)}
\end{equation}

Encryption reduces the transfer performance down to about 90\% compared with the transfer rates of non encrypted data transfers. For smaller message sizes this negative effect of encryption gets even more and more negligible. In other words, especially for small message sizes encryption is not a substantial performance killer compared to SDVN impact (see Figure \ref{fig:relcomparison}b). For bigger message sizes the data transfer performance impact of encryption is comparable to containerization.

\begin{table}[tb]
\caption{\uppercase{Relative standard deviations of measured transfer rates}}
\label{tab:characteristics} 
\begin{tabular}{lr|ccc}
 & & & \footnotesize \textbf{\%RSD} & \\
\footnotesize \textbf{Experiment} & \footnotesize \textbf{Data} & \footnotesize \textbf{Min} & \footnotesize \textbf{Avg} & \footnotesize \textbf{Max} \\
  \hline
\small Reference   & \footnotesize 90 GB                  & \footnotesize 0,9  & \footnotesize 2,9 & \footnotesize 15,9 \\
\small Cross Regional & \footnotesize 19 GB            & \footnotesize 0,9  & \footnotesize 14,9 & \footnotesize 28,7 \\
\small Docker        & \footnotesize 57 GB                 & \footnotesize 0,8 & \footnotesize 2,0 & \footnotesize 10,3 \\
\small Docker\_SDVN & \footnotesize 75 GB & \footnotesize 0,4 & \footnotesize 11,3 & \footnotesize 21,2 \\
\small Docker\_Encrypted & \footnotesize 75 GB & \footnotesize 0,5 & \footnotesize 10,9 & \footnotesize 16,2 \\
\end{tabular}
\end{table}

\section{\uppercase{Conclusion}}
\label{sec:conclusion}

  This study analyzed performance impact of containers, overlay networks and encryption to overall network performance of HTTP-based and REST-like services deployed to IaaS cloud infrastructures. Obviously, our conclusions should be cross checked with other SDVN solutions for containers. The provided data \cite{sdvn-impact-database} and benchmarking tools to apply the presented methodology \cite{ping-pong} can be used as benchmark for that purpose. Nevertheless, some of the study results can be used to derive some design recommendations for cloud deployed HTTP-based and REST-like systems of general applicability.

Although \textbf{containers} are stated to be lightweight \cite{SPF+2007}\cite{FFRR2014}, this study shows that container impact on network performance is not negligible. Containers show a performance impact of about 10\% to 20\%. The impact of \textbf{overlay networks} can be even worse. The analyzed SDVN solution showed a performance impact of about 30\% to 70\%, which is comparable to a cross regional deployment of a service between Ireland and Japan. \textbf{Encryption} performance loss is minor, especially for small message sizes.

The results show that performance impacts of overlay networks can be minimized on several layers. On \textbf{application layer} message sizes between system components should be minimized whenever possible. Network performance impact gets worse with increasing message sizes. On \textbf{overlay network layer} performance could be optimized by 10\% to 20\% by providing SDVN router applications directly on the host (in a not containerized form, because 10\% to 20\% are due to general container losses). On \textbf{infrastructure layer}, the SDVN routers should be deployed to multi core virtual machines to avoid situations, where SDVN routers contend for CPU with application processes.

So containers, which are often mentioned to be lightweight, are not lightweight under all circumstances. Nevertheless, the reader should not conclude to avoid container and SDVN technologies in general. Container and SDVN technologies provide more flexibility and manageability in designing complex horizontally scalable distributed cloud systems. And there is nothing wrong about flexibility and manageability of complex systems. That is why container solutions like Docker and microservice approaches regain so much attention recently. But container and SDVN technologies should be always used with above mentioned performance implications in mind.

\section*{Acknowledgment}
This study was funded by German Federal Ministry of Education and Research (Project Cloud TRANSIT, 03FH021PX4). The author thanks Lübeck University (Institute of Telematics) and fat IT solution GmbH (Kiel) for their support of Cloud TRANSIT. The author also thanks Bryan Boreham of zett.io for checking our data of zett.io's \emph{weave} solution (which might show now better results than the analyzed first version of \emph{weave}).

\bibliographystyle{IEEEtran}
\bibliography{references}

\end{document}